\documentclass[twocolumn,floatfix,nofootinbib,aps,reprint]{revtex4}
\usepackage{times}
\usepackage{nicefrac}
\usepackage{latexsym}
\usepackage{amsmath, mathrsfs, bm}
\usepackage{eqnarray}
\usepackage{url}
\usepackage{cases}
\usepackage{epsfig}
\usepackage{color}
\usepackage{graphicx,subfigure}
\usepackage{grffile}
\usepackage{float}
\usepackage{natbib}
\usepackage{soul}

\newcolumntype{L}[1]{>{\raggedright\arraybackslash}p{#1}}
\newcolumntype{C}[1]{>{\centering\arraybackslash}p{#1}}  
\newcolumntype{R}[1]{>{\raggedleft\arraybackslash}p{#1}}

\def\hi{\textsc{Hi~}}

\def\cambs{\textsc{Camb\_Sources~}}

\def\flocal{{f^{\rm local}_{\rm NL}}}

\begin{document}

\title{{\textbf Constraints on primordial non-Gaussianity from future \hi 
intensity mapping experiments}}
\author{\textsc{Yi-Chao Li~}}
\email{lixiating@gmail.com}
\author{\textsc{Yin-Zhe Ma~}}
\email{ma@ukzn.ac.za}
\affiliation{School of Chemistry and Physics, University of KwaZulu-Natal, Westville Campus, Private Bag X54001, Durban, 4000, South Africa}
\affiliation{NAOC-–UKZN Computational Astrophysics Centre (NUCAC), University of KwaZulu-Natal, Durban, 4000, South Africa}

\begin{abstract}
The primordial non-Gaussianity induces scale-dependent bias of the \hi with respect to the underlying dark matter, which exhibits features on the very large scales of the 21-cm power spectrum potentially observable with \hi intensity mapping observations. We forecast the prospective constraints on the four fundamental shapes of primordial non-Gaussianity (local, equilateral, orthogonal, and enfolded), with the current and future \hi intensity mapping experiments, BINGO, FAST, and SKA-I. With the current configuration of the experiments and assumed one-year observation time, we find that the SKA-I will provide tighter constraints on the local shape of primoridal non-Gaussianity than {\it Planck}. The results are
$(\sigma_{f^{\rm local}_{\rm NL}},\sigma_{f^{\rm equil}_{\rm NL}},\sigma_{f^{\rm orth}_{\rm NL}},\sigma_{f^{\rm enfold}_{\rm NL}})_{\rm SKA-I}=(0.54, 86, 25, 43)$, $(\sigma_{f^{\rm local}_{\rm NL}},\sigma_{f^{\rm equil}_{\rm NL}},\sigma_{f^{\rm orth}_{\rm NL}},\sigma_{f^{\rm enfold}_{\rm NL}})_{\rm BINGO}=(17, 100, 128, 164)$, $(\sigma_{f^{\rm local}_{\rm NL}},\sigma_{f^{\rm equil}_{\rm NL}},\sigma_{f^{\rm orth}_{\rm NL}},\sigma_{f^{\rm enfold}_{\rm NL}})_{\rm FAST}=(9.5, 44, 75, 94)$. If the lower frequency band of FAST can be used, the constraint on 
    local-type primordial non-Gaussianity will be $\sigma_{f_\mathrm{NL}}\sim1.62$ which is better than {\it Planck}. In addition, if the observation time for FAST could be extended to two years, 
    the constraint on the equilateral shape of primordial non-Gaussianity would
    be improved to $\sigma_{f_\mathrm{NL}}\sim32$. Similarly, if the observational time of SKA-I could be
    extended to two years, the constraint on local and orthogonal shapes could be improved to $0.43$ and $20$, respectively, achieving better constraints than {\it Planck}. 
\end{abstract}

\maketitle

\section{introduction}
\label{sec:intro}
The statistical properties of the primordial fluctuation offer 
rich insights into the physics of inflation and the early Universe
~\cite{2004PhR...402..103B}.
One of the widely discussed questions is whether or not the 
primordial fluctuations deviated from the Gaussian distribution.
The simple single-field slow-roll inflationary model predicts 
primordial fluctuation with almost Gaussian distribution
~\cite{1992PhRvD..46.4232F,2003NuPhB.667..119A,2003JHEP...05..013M}.
However, many alternative models of single-field slow-roll inflation can
produce different types of primordial non-Gaussianity
~\cite{1997PhRvD..56..535L,2003PhRvD..67b3503L,
2004PhRvD..70l3505A, 2006JCAP...05..004C,2007JCAP...01..002C, 
2008JCAP...05..001H,2009JCAP...05..018M,2011JCAP...01..030A,
2011JCAP...03..025A,2014JCAP...02..025A} (PNG), 
which leaves distinctive features in the statistical properties of the 
cosmic microwave background (CMB)
and the large-scale structure (LSS) of the Universe.

If the primordial fluctuation is Gaussian, the two-point correlation function (i.e. the power spectrum in Fourier space) can describe all of the statistical properties of the PNG. Therefore, the most straightforward way to measure the PNG is through the higher-order correlation of CMB or LSS. Current measurements of the temperature and polarization of CMB from the {\it Planck} satellite provide state-of-the-art constraints on local, equilateral and orthogonal types of PNG~\cite{2016A&A...594A..17P} as $f_{\rm NL}^{\rm local}=0.8 \pm 5.0$, $f_{\rm NL}^{\rm equil}=-4 \pm 43$, and $f_{\rm NL}^{\rm ortho}=-26 \pm 21$ at $68\%$ confidence level (C.L.).

Besides the constraints from CMB, there have been many efforts to measure 
$f_{\rm NL}$ through large-scale structure surveys. This is because the PNG 
induces a scale-dependent 
bias of the galaxy with respect to the underlying dark matter distribution 
tracer ~\cite{2008PhRvD..77l3514D,2008ApJ...677L..77M,2008ApJ...684L...1C, 
2008JCAP...08..031S, 2011PhRvD..84f3512D, 2012PhRvD..86f3518M,
2015arXiv150705903R}. 
Reference~\cite{2008JCAP...08..031S} used spectroscopic and photometric luminous 
red galaxy samples and quasar samples from the SDSS survey to obtain the limit 
for local-type PNG as $-31(-96)< f^{\rm local}_{\rm NL}<+70(+96)$ at 
$95\%$ ($99.7\%$) C.L., which was comparable to the measurements at the time from the
{\it Wilkinson Microwave Anisotropy Probe} ({\it WMAP}) five-year results. 
Reference~\cite{Xia11} used radio sources from the NRAO VLA Sky Survey (NVSS), 
the quasar and MegaZ-LRG (DR7) catalogues of
the SDSS, and the final SDSS II Luminous Red Galaxy (LRG) photometric 
redshift survey and found $f^{\rm local}_{\rm NL}=48 \pm 20$ ($1\sigma$ C.L.). 
Reference~\cite{Nikoloudakis13} found $f^{\rm local}_{\rm NL} = 90 \pm 30$ at 
$1\sigma$ C.L. by using photometric SDSS data, but due to unaccounted systematics 
this result may be better interpreted as $f^{\rm local}_{\rm NL} < 120$ 
at $84\%$ C.L. Reference~\cite{Ross13} used the 
SDSS-III Baryon Oscillation Spectroscopic Survey (BOSS)
data to constrain the $\flocal$ and found $-45<\flocal < 195$ at $2\sigma$ C.L. 
In addition, Ref.~\cite{2013MNRAS.436.2029M} used the correlation of 
the residual peculiar velocities on different directions to constrain 
PNG and found $|\flocal|<25.7$ at $68\%$ C.L. 
These limits are currently consistent with but weaker than the measurements 
from the {\it Planck} CMB observation.  
In general, the scale-dependent bias signal can be degenerated with 
the nonlinear bias between halo and underlying dark matter, which is
contributed from the nonlinear evolution of the matter fluctuations
~\cite{2016arXiv161206366G}.
However, forecasts indicate that the constraint errors could decrease 
$1$ or $2$ orders of the magnitude with the future LSS survey, especially 
for the future radio survey. (see \cite{2016arXiv161109787D} and its 
references for review). 
Beside the constraint on the PNG amplitude, previous studies also
show that the scale-dependent bias introduced by the PNG is sensitive 
mainly to the squeezed limit and, with the future LSS surveys, 
it can be used to distinguish among different PNG shapes
~\cite{2012JCAP...08..019N,2012JCAP...08..033S}.

The scale-dependent bias not only affects the 
large-scale galaxy bias, but also affects the \hi distribution. A more efficient method of the radio survey is to map out a large volume of the Universe through the intensity mapping technique, which measures the combined \hi emission of
the unresolved galaxies. Therefore, in principle one can obtain a three-dimensional \hi distribution that can provide more modes of fluctuation than the CMB two-dimensional sphere. There have been several works to forecast the detectability of PNG through the \hi intensity mapping technique~\cite{2013PhRvL.111q1302C, 2015ApJ...798...40X,Bull15}, but those forecasts are exclusively only for the local and equilateral type of PNG and limited experimental cases (SKA and Tianlai). In this work, we will calculate the scale-dependent bias of all four typical types of PNG by using the halo model and calculate their imprints on the power spectrum of \hi. Then we forecast the detectability of all three ongoing \hi imaging surveys, i.e. BAO as Integrated Neutral Gas Observation (BINGO)~\cite{2013MNRAS.434.1239B}, Five-Hundred-Metre Aperical Spherical Telescope (FAST)~\cite{2011IJMPD..20..989N,2016RaSc...51.1060L} and Square Kilometre Array Phase-I (SKA-I)~\cite{2009IEEEP..97.1482D}.

This paper is organized as follows. In Sec.~\ref{sec:power-bispec}, we 
summarize the primordial bispectrum and 
discuss different types of PNG to be forecasted in this work. 
In Sec.~\ref{sec:bhi}, we calculate the 
scale-dependent bias of the LSS induced by the PNG, and then the power 
spectrum of \hi. In Sec.~\ref{sec:fisher}, we introduce the Fisher matrix 
forecast method that used in our analysis. In Sec.\ref{sec:exp}, 
the detailed experiment
parameters are discussed. In Sec.~\ref{sec:result}, we present 
our results and some discussion. Conclusion will be in the last section.

Besides the PNG parameters, we will adopt a spatially flat Universe with cosmological parameters fixed as {\it Planck} 2015 best-fitting values~\cite{Planck2015-parameter}, i.e. $\Omega_{\rm m}=0.309$; $\Omega_{\Lambda} = 0.691$; $\sigma_{8} = 0.809$; and $h = 0.68$, where the Hubble constant is $H_{0} = 100\,h\,{\rm km}\,{\rm s}^{-1}\,{\rm Mpc}^{-1}$. The amplitude and tilt of scalar power spectrum are $A_{\rm s}(k_{0})=2.141 \times 10^{-9}$ and $n_{\rm s}=0.961$, where pivot scale is $k_{0}=0.002\,{\rm Mpc}^{-1}$.

\section{Primordial bispectrum}
\label{sec:power-bispec}


The inflationary models predict the primordial 
curvature fluctuations with the deviation from 
Gaussian distribution~\cite{1987PhLB..197...66A,1994ApJ...430..447G,2003JHEP...05..013M,2015CRPhy..16..969R}. 
The deviation is particularly described by writing the gauge-invariant
Bardeen's potential $\phi$ as the sum of a Gaussian random field 
and a quadratic correlation~\cite{1994ApJ...430..447G,2001PhRvD..63f3002K},
\begin{equation}
    \phi = \phi_\mathrm{G} + f_\mathrm{NL}(\phi_\mathrm{G}^2 - 
    \langle \phi_\mathrm{G}^2 \rangle),
\end{equation}
in which $f_\mathrm{NL}$ is a dimensionless, phenomenological parameter
describing the magnitude of the PNG.

To extract more information of the non-Gaussian primordial
fluctuations, we need to go beyond the statistics of the power spectrum.
The lowest-order statistics sensitive to 
the PNG is the three-point function or
bispectrum $B_\phi(k_1, k_2, k_3)$, in which $\phi$ is the 
primordial Bardeen potential which is directly related to the curvature 
perturbation~\cite{1980PhRvD..22.1882B}. 
The potential of the primordial curvature perturbation 
is related to the Newtonian potential during the matter domination 
via the transfer function $T(k)$ which satisfies $T(k\rightarrow0)=1$.
By applying the Poisson equation, $\phi$ is related to the matter
density field $\delta_{\rm m}(k)$ by $\delta_{\rm m}(k) = \mathcal{M}(k)\phi(k)$,
where
\begin{equation}\label{eq:m}
    \mathcal{M}(k) = \frac{2}{3}\frac{k^2T(k)}{\Omega_{\rm m}H_0^2}.
\end{equation}
The configuration shape of $B_\phi(k_1, k_2, k_3)$ is related to 
the physical mechanisms during the inflation. In our analysis,
we consider four classes of bispectrum shape characterizing
the local, equilateral, enfolded and orthogonal types of PNG.

\subsection{Local shape}
The local-type PNG can be produced in
different inflationary models, such as the multifield model
\cite{1997PhRvD..56..535L,2006JCAP...01..006A}, 
curvaton model~\cite{2003PhRvD..67b3503L},
inhomogeneous reheating~\cite{2004PhRvD..69b3505D} 
or new Ekpyrotic models~\cite{2008PhRvD..77f3533L}.
In these cases, $f_\mathrm{NL}^\mathrm{local}$ can be substantially 
different from zero.

The potential bispectrum of the local-type PNG 
has the simple form,
\begin{equation}
    \begin{split}
        B_\phi(k_1, k_2, k_3) = 2 &f_\mathrm{NL}^\mathrm{local} 
        \left[ P_\phi(k_1)P_\phi(k_2) + (\mathrm{cyc.})\right], \label{eq:B-local}
    \end{split}
\end{equation}
in which, 
$P_\phi(k)=2\pi^2 A_{\rm s}(k_0) (k/k_0)^{n_{\rm s}-4}$
is the power spectrum of the Gaussian Bardeen potential. 

\subsection{Equilateral shape}
The equilateral-type of PNG can be produced in the inflationary
models with higher-derivative interactions. 
Usually there are two dominant interaction terms of the inflation
field giving rise to the PNG peaking in the equilateral limit, which 
can be represented by a unique template with the equilateral shape.

The primordial bispectrum 
of the equilateral type takes the form \cite{2006JCAP...05..004C},
\begin{equation}
    \begin{split}
        B_\phi(k_1, k_2, k_3) &= 6 f_\mathrm{NL}^\mathrm{equil}
        \gamma(k_1, k_2, k_3)\vphantom{P^{2/3}}\\
        &\times\left[\vphantom{P^{2/3}}
        -\left(\vphantom{P^{2/3}}
        P_\phi(k_1)P_\phi(k_2) + (\mathrm{cyc.})\right)
        \right.\\
        &\left. 
        -\,2\left(\vphantom{P^{2/3}}
        P_\phi(k_1)P_\phi(k_2)P_\phi(k_3)\right)^{2/3}
        \vphantom{P^{2/3}}\right. \\
        &\left.
        + \left(P_\phi^{1/3}(k_1)P_\phi^{2/3}(k_2)P_\phi(k_3) + 
        (\mathrm{cyc.}) \right)
        \vphantom{P^{2/3}}\right], \label{eq:B-equi}
    \end{split}
\end{equation}
in which function $\gamma(k_1, k_2, k_3)$ takes into account the running of $f_\mathrm{NL}^\mathrm{equil}$ and reads
~\cite{2008JCAP...04..014L},
\begin{equation}
    \gamma(k_1, k_2, k_3) = \left[\frac{k_1 + k_2 + k_3}{k_\mathrm{CMB}}
    \right]^{-2\kappa},
\end{equation}
where $k_\mathrm{CMB}=0.086\,h\mathrm{Mpc}^{-1}$, roughly 
corresponding to the largest $\ell$ used to estimate the non-Gaussianity
with {\it WMAP} data~\cite{2009ApJS..180..330K}.
The free parameter $\kappa$ is assumed to be constant. Following 
the discussion in the works of
\citep{2008JCAP...04..014L,2009MNRAS.394..133C}, 
we use small negative $\kappa=-0.2$ to enhance 
the non-Gaussianity on small scales. In the rest of this paper, 
the equilateral-type bispectrum always take the form of Eq.~(\ref{eq:B-equi})
with $\kappa=-0.2$.

\subsection{Orthogonal shape}
The shapes of PNG caused by the two dominant terms of higher-derivative 
interactions, as we introduced above, are slightly different around 
flattened triangles $k_2+k_3\simeq k_1$. By taking an appropriate 
linear combination, the resulting orthogonal shape of the PNG can minimize the 
similarities and maximize the differences. The orthogonal shape
is well approximated by the following template
~\cite{2015CRPhy..16..969R,2010JCAP...01..028S}:

\begin{equation}
    \begin{split}
        B_\phi(k_1, k_2, k_3) &= 6 f_\mathrm{NL}^\mathrm{orth}
        \left[\vphantom{P^{2/3}}
        -3\left(\vphantom{P^{2/3}}
        P_\phi(k_1)P_\phi(k_2) + (\mathrm{cyc.})\right)
        \right.\\
        &\left. 
        -\,8\left(\vphantom{P^{2/3}}
        P_\phi(k_1)P_\phi(k_2)P_\phi(k_3)\right)^{2/3}
        \vphantom{P^{2/3}}\right. \\
        &\left.
        +\,3\left(P_\phi^{1/3}(k_1)P_\phi^{2/3}(k_2)P_\phi(k_3) + 
        (\mathrm{cyc.}) \right)
        \vphantom{P^{2/3}}\right], \label{eq:B-orth}
    \end{split}
\end{equation}

\subsection{Enfolded shape}
It is well studied that if the initial vacuum state for the inflation 
deviates from the standard Bunch-Davies vacuum, the resulting bispectrum 
takes the enfolded shape
~\cite{2007JCAP...01..002C,2008JCAP...05..001H,
2009JCAP...05..018M,2011JCAP...01..030A},
which can be approximated by
\begin{equation}
    \begin{split}
        B_\phi(k_1, k_2, k_3) &= 6 f_\mathrm{NL}^\mathrm{enfold}
        \left[\vphantom{P^{2/3}}
        \left(\vphantom{P^{2/3}}
        P_\phi(k_1)P_\phi(k_2) + (\mathrm{cyc.})\right)
        \right.\\
        &\left. 
        +\,3\left(\vphantom{P^{2/3}}
        P_\phi(k_1)P_\phi(k_2)P_\phi(k_3)\right)^{2/3}
        \vphantom{P^{2/3}}\right. \\
        &\left.
        -\left(P_\phi^{1/3}(k_1)P_\phi^{2/3}(k_2)P_\phi(k_3) + 
        (\mathrm{cyc.}) \right)
        \vphantom{P^{2/3}}\right]. \label{eq:B-enfold}
    \end{split}
\end{equation}
Note that as pointed out in Appendix C of~\cite{Creminelli11}, the squeezed limit of this type of non-Gaussianity will result in a negligible scale-dependent bias. Reference~\cite{Creminelli11} suggested a new factorizable template with correct squeezed limit.

\section{\hi bias and power spectra of 21-cm}
\label{sec:bhi}
The \hi bias is the bias of \hi distribution with respect to the 
underlying dark matter distribution
and the \hi bias function, $b_\mathrm{\hi}$, can be obtained by  assuming 
a model for the amount of \hi mass in a dark matter halo of mass $M$,
$M_\mathrm{\hi}(M)$, and integrating 
over the halo mass function $\mathrm{d}n/\mathrm{d}M$. Here we use the Sheth-Tormen halo mass function~\cite{Sheth02} with mass range [$10^{8}$, $10^{16}$]\,${\rm M}_{\odot}$
\begin{equation}\label{eq:b}
    b_\mathrm{\hi}(z) = \frac{1}{\rho_\mathrm{\hi}(z)} 
    \int^{M_\mathrm{max}}_{M_\mathrm{min}} \mathrm{d}M 
    \frac{\mathrm{d}n}{\mathrm{d}M}(M, z)
    M_\mathrm{\hi}(M) b(M, z),
\end{equation}
in which $b(M, z)$ is the real-space halo bias and $\rho_\mathrm{\hi}(z)$ is,
\begin{equation}\label{eq:rho}
    \rho_\mathrm{\hi}(z) = 
    \int^{M_\mathrm{max}}_{M_\mathrm{min}} \mathrm{d}M 
    \frac{\mathrm{d}n}{\mathrm{d}M}(M, z)
    M_\mathrm{\hi}(M) .
\end{equation}
For the \hi intensity mapping experiments, we follow the assumption 
discussed in~\cite{2015aska.confE..19S} and consider a simple power 
law model for the amount of \hi mass,
\begin{equation}
    M_\mathrm{\hi}(M) = AM^\alpha, \, \alpha \simeq 0.6,
\end{equation}
which is a redshift independent function.
The prefactor $A$ will be canceled with the normalization of 
$\rho_\mathrm{\hi}$.

\subsection{The Lagrangian bias}

\begin{figure}[htp]
    \centering
    \includegraphics[width=0.5\textwidth]{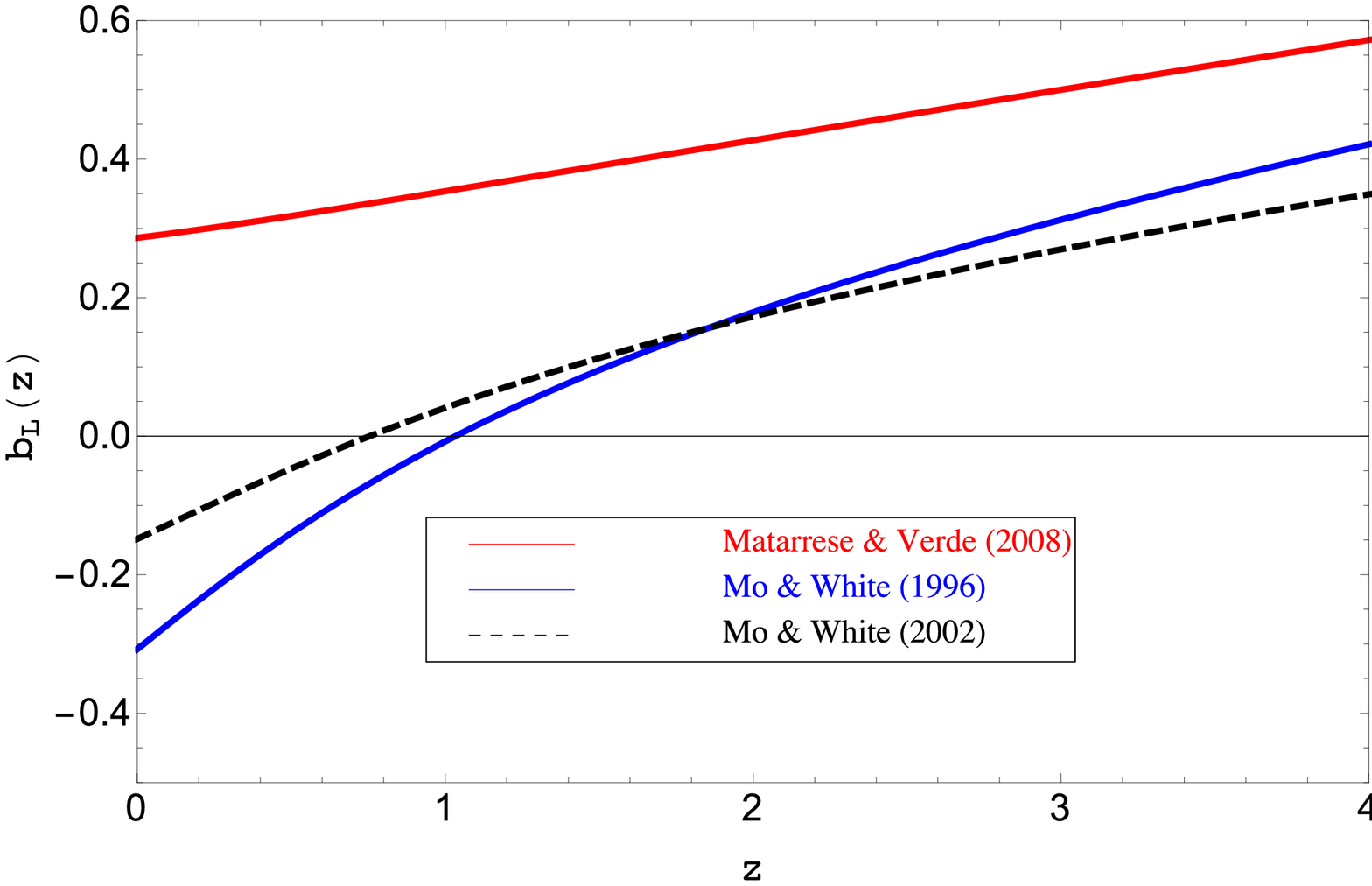}
    \vspace{-0.6cm}
    \caption{Three models of Lagrangian bias $b_{\rm L}(z)$, i.e.,~\citet{2008ApJ...677L..77M},~\citet{1996MNRAS.282..347M}, and~\citet{2002MNRAS.336..112M}.}\label{fig:bhi}
\end{figure}
The Lagrangian bias describes the statistical bias of the halo distribution 
to the primordial dark matter fields. 
The PNG affects the initial conditions of
the primordial density fields, so it is more convenient to study such 
effects in Lagrangian space. 
On the other hand, it is also necessary to study the statistics of the 
evolved halo field at low redshifts in Eulerian space, which is conveniently
related to the observation.
The bias in Lagrangian space, $b_\mathrm{L}$,
relates to the Eulerian space bias, $b_\mathrm{E}$, via 
$b_\mathrm{E} = b_\mathrm{L} + 1$~\cite{1996MNRAS.282..347M}.
The extra unity factor of $b_\mathrm{E}$ reflects the motions of 
primordial peaks at later times~\cite{2013MNRAS.436.2029M}. 
The uniformly distributed halos in the initial epoch,
which have $b_\mathrm{L}=0$, will lead to unbiased distribution 
to the dark matter field at a later time. 
The $b_\mathrm{L}$ for halos is defined as positive. But for other
dark matter tracers, it can be negative. The tracers anticorrelated 
with the initial dark matter fields will lead to the less clustered
distribution than the dark matter field at later time.

It the past $30$ years, people have 
been developing different analytical, semianalytical and parametric models 
of the bias function. Below, we list the three most typical and 
commonly used ones.

Based on the Press and Schechter (PS) halo mass function
~\cite{1974ApJ...187..425P} and its extensions, 
\citeauthor{1996MNRAS.282..347M}~(\citeyear{1996MNRAS.282..347M}) 
\citep{1996MNRAS.282..347M} give the bias factor for the halo
of mass $M$,
\begin{equation}
    b_\mathrm{L}(M,z) = \frac{1}{\delta_{\rm c}}\left[\nu^2(M,z)-1\right],
\end{equation}
where $\nu(M, z)=\delta_{\rm c}(z)/\sigma_R$. $\delta_{\rm c}(z)=\delta_{\rm c}/D(z)$,
where $D(z)$ is the linear growth function and we use Eq.~(10) in~\cite{2002MNRAS.336..112M} to compute it. $\delta_{\rm c}\simeq 1.686$ is the critical density
contrast for spherical collapse. With the approximation of high-peak, the above bias factor can be
expressed as $b_\mathrm{L}(M, z)=\delta_{\rm c}(z)/\sigma_R^2$
(\citeauthor{2008ApJ...677L..77M}~\citeyear{2008ApJ...677L..77M}
\citep{2008ApJ...677L..77M}).
With the ellipsoidal collapse model~\cite{2001MNRAS.323....1S},
\citeauthor{2002MNRAS.336..112M}~(\citeyear{2002MNRAS.336..112M})
\citep{2002MNRAS.336..112M} give 
another expression,
\begin{eqnarray}
 b_\mathrm{L}(M,z) &=& \frac{1}{\delta_{\rm c}(z)}
        \left[\nu'^2 + b\nu'^{2(1-c)} \right. \nonumber \\
        &-& \left.\frac{\nu'^{2c}/\sqrt{a}}{\nu'^{2c}+b(1-c)(1-c/2)}\right],
        \label{eq:bl}
\end{eqnarray}
in which, $\nu'=\sqrt{a}\nu$ and $a=0.707,\,b=0.5,\,c=0.6$.

Figure~\ref{fig:bhi} shows the three models of Lagrangian bias we discussed above.

\subsection{The scale-dependent bias}
\label{sec:scale-dependent}
\begin{figure*}[htb]
    \centerline{
    \includegraphics[width=0.5\textwidth]{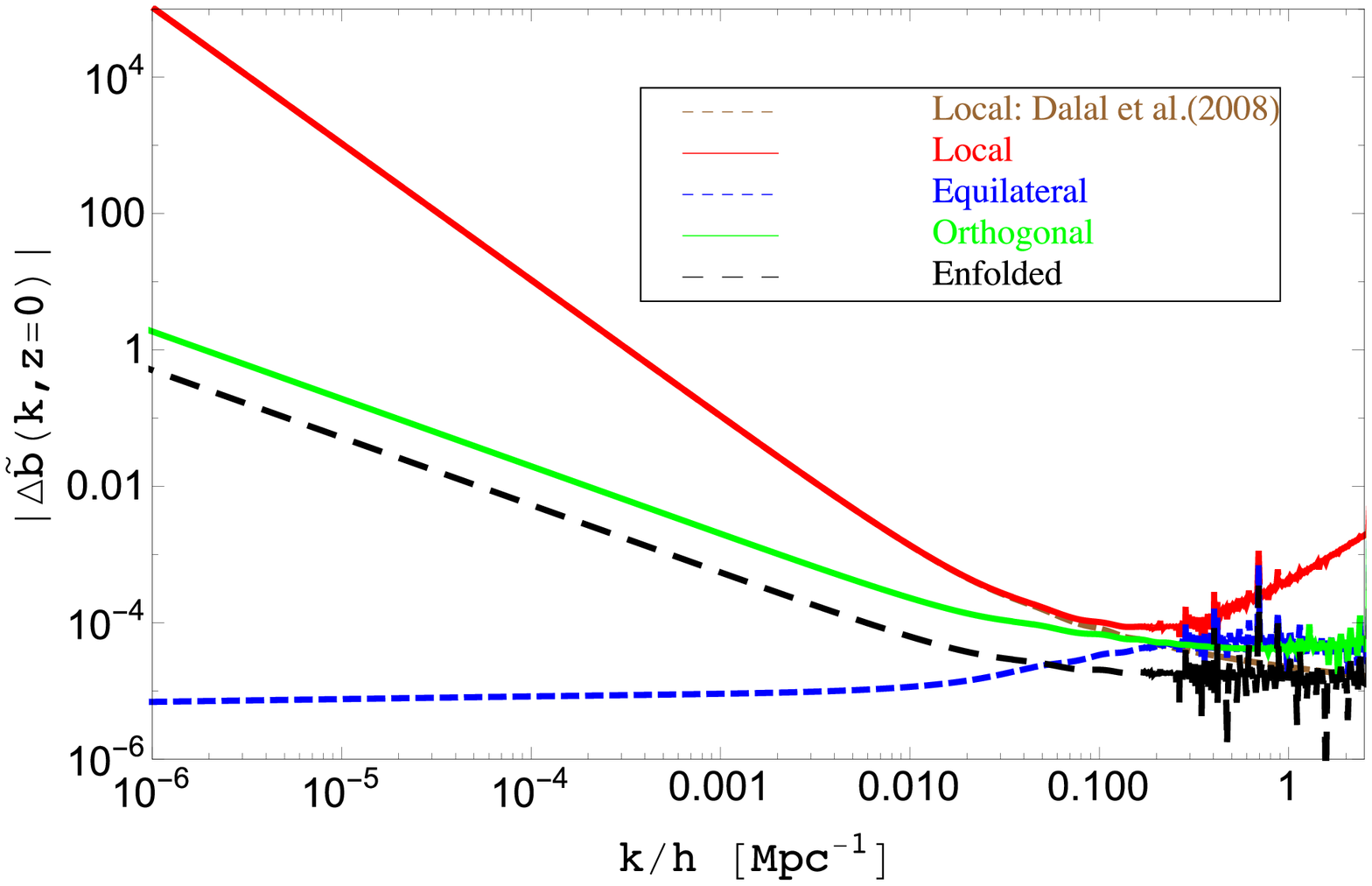}
    \includegraphics[width=0.5\textwidth]{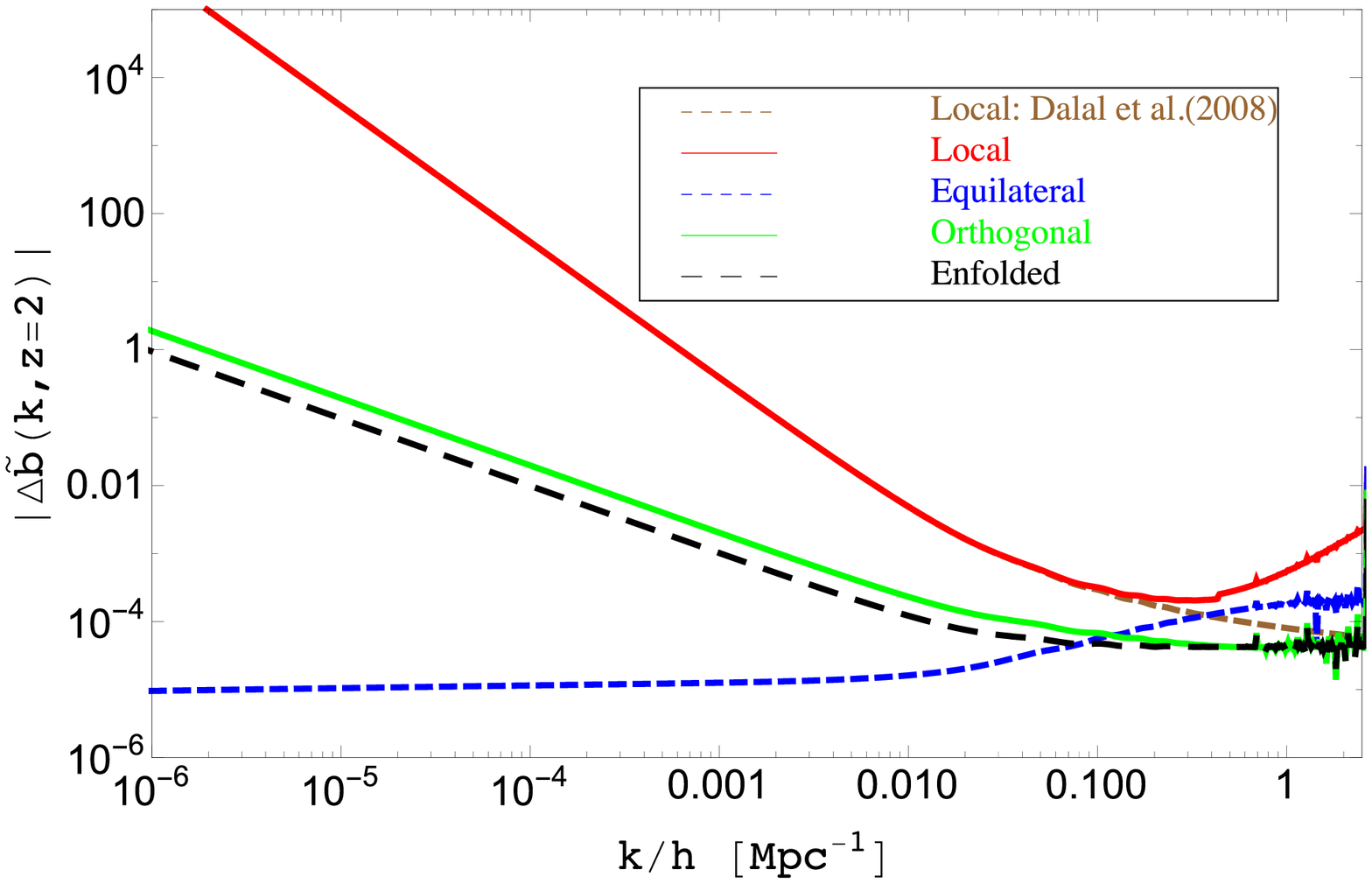}}
    \caption{The absolute value of scale-dependent bias $|\Delta b(z,k)|$ [Eq.~(\ref{eq:Delta-b-z-k})] for different PNG shapes at $z=0$ (left panel) and $z=2$ (right panel) with assumed $f_{\rm NL}=1$. The four shapes of PNG are shown in different colors and dashed lines listed in the legend. The reason to plot the absolute value is because the orthogonal shape of $\Delta b$ is negative (see also Fig.~1 in~\cite{Fedeli11}). The approximation of the local shape of PNG by~\citet{2008PhRvD..77l3514D} [Eq.~(\ref{eq:db_dalal})] is shown in the brown dashed line, which is consistent and almost completely overlapped with the computation from the halo model [Eqs.~(\ref{eq:B-local}) and (\ref{eq:db_mv})] shown with the red solid line.}\label{fig:db}
\end{figure*}

As we analyzed before, PNG affects the distribution of the peaks at the initial
stage of matter fluctuations; therefore, it is correlated with the Lagrangian bias.
In the presence of PNG, the halo bias can be written 
as the combination of a usual scale-invariant bias, $b(M, z)$, 
and a scale-dependent modification, $\Delta b(M, z, k)$,
\begin{equation}\label{eq:b2}
    b^\mathrm{NG}(M, z, k) = b(M, z) + \Delta b(M, z, k).
\end{equation}
By substituting Eq.~(\ref{eq:b2}) into Eq.~(\ref{eq:b}),
we can obtain the scale-dependent \hi bias, which can be
expressed as,
\begin{equation}\label{eq:bNG}
    b_\mathrm{\hi}^\mathrm{NG}(z, k) = b_\mathrm{\hi}(z) + 
    \Delta b_\mathrm{\hi}(z, k),
\end{equation}
in which $b_\mathrm{\hi}^\mathrm{NG}(z, k)$ is the total bias, $b_\mathrm{\hi}(z)$ is the scale-independent term, and $\Delta b_\mathrm{\hi}(z, k)$ is the scale-dependent term, which is obtained by integrating
$\Delta b(M,z,k)$ over the 
halo mass function and the \hi mass model,
\begin{equation}
    \begin{split}
        \Delta b_\mathrm{\hi}(z, k) &= \frac{1}{\rho_\mathrm{\hi}(z)}
        \int^{M_\mathrm{max}}_{M_\mathrm{min}} \mathrm{d}M \\
        &\times
        \frac{\mathrm{d}n}{\mathrm{d}M}(M, z)
        M_\mathrm{\hi}(M) \Delta b(M, z, k), \label{eq:Delta-b-z-k}
    \end{split}
\end{equation}
where $\rho_\mathrm{\hi}(z)$ is calculated in Eq.~(\ref{eq:rho}).

\citet{2008PhRvD..77l3514D} firstly derived the expression of
scale-dependent correction to the bias of galaxies and halos 
for local-shape bispectrum,
\begin{equation}\label{eq:db_dalal}
    \Delta b^\mathrm{D}(z, k) = 2 (b_\mathrm{E}-1)f_\mathrm{NL}
    \delta_{\rm c} \frac{3\Omega_{\rm m}}{2 a(z) g(z) r_H^2 k^2},
\end{equation}
in which, $\delta_{\rm c}$ is the critical density, $a(z) g(z)=D(z)$
is the linear growth factor and $r_H=1/H_0$.  
Equation~(\ref{eq:db_dalal}) is derived 
by only considering the high peaks of the density contrast, 
which means that the expression only works at the large scales
with $k \rightarrow 0$.

More accurate analytical expressions for the scale-dependent bias
have been studied~\cite{2008ApJ...677L..77M,2008ApJ...684L...1C, 
2008JCAP...08..031S, 2011PhRvD..84f3512D, 2012PhRvD..86f3518M,
2015arXiv150705903R}. A widely used expression is derived by~\citet{2008ApJ...677L..77M},
\begin{equation}\label{eq:db_mv}
    \Delta b^\mathrm{MV}(M, z, k) = 2 f_\mathrm{NL}
    \left(\frac{\delta_{\rm c}^{2}(z)}{\sigma^2_R}\right)
     \frac{\mathcal{F}(k)}{\mathcal{M}_R(k)},
\end{equation}
in which, $\delta_{\rm c}(z)=\delta_{\rm c}/D(z)$~\footnote{This is consistent with
Eq.~13 in~\cite{2008ApJ...677L..77M}. The ``$\Delta_{\rm c}$'' defined in~\cite{2008ApJ...677L..77M}
is equal to $\delta_{\rm c}$ in this paper.} and $\mathcal{M}_R(k)$ is Eq.~(\ref{eq:m}) smoothed with window function $W_R(k)$,
\begin{equation}
    \mathcal{M}_R(k) = \frac{2}{3}\frac{T(k)k^2}{H^2_0\Omega_{\rm m}}W_R(k),
\end{equation}
where 
$R$ denotes a smoothing radius which defines the halo mass $M$ by
\begin{equation}\label{eq:sigmaR}
    M=\frac{3H_0^2\Omega_{\rm m}}{8\pi G}\frac{4\pi R^3}{3}.
\end{equation}
So $\Delta b^\mathrm{MV}$
is also a function of halo mass, $M$. $\mathcal{F}(k)$ is related 
to the bispectrum of primordial potential field $B_\phi(k_1, k_2, k)$,
and the power spectrum $P_\phi(k)$,
\begin{equation}\label{eq:fk}
    \begin{split}
        \mathcal{F}(k) = \frac{1}{16\pi^2\sigma_R^2} &
        \int \mathrm{d}k_1 k_1^2 \mathcal{M}_R(k_1) \\
        & \times \int_{-1}^{1} \mathrm{d}\mu \mathcal{M}_R(k_2)
        \frac{B_\phi(k_1, k_2, k)}{P_\phi(k)},
    \end{split}
\end{equation}
where $k_2^2 = k^2 + k_1^2 + 2kk_1\mu$ and $\sigma_R$ is the {\it rms}
of the underlying dark matter fluctuation fields smoothed on scale
$R$ given in Eq.~(\ref{eq:sigmaR}).

If we substitute the local-shape bispectrum into Eq.~(\ref{eq:fk}),
and take the limit of $k\rightarrow0$, then\footnote{In Ref.~\cite{2008ApJ...677L..77M}, $b_{\rm E}-1=b_{\rm L}=\delta_{\rm c}/\sigma^{2}_{R}$} the dependence of $\Delta b^{\rm MV}(M,z,k)$ on the halo mass
automatically drops of,
\begin{eqnarray}
    \mathcal{F}(k\rightarrow0) &\rightarrow& 1 \nonumber\\
    T(k\rightarrow0)             &\rightarrow& 1 \nonumber\\
    \mathcal{M}_R(k\rightarrow0) &\rightarrow& (2/3)k^2/(H_0^2\Omega_{\rm m}),\nonumber
\end{eqnarray}
and,
\begin{equation}
    \begin{split}
        \Delta b^\mathrm{MV}(z, k\rightarrow0) & \rightarrow
        2 (b_E-1) f_\mathrm{NL} \frac{\delta_{\rm c}}{a(z)g(z)} 
        \frac{3}{2}\frac{H_0^2\Omega_{\rm m}}{k^2} \\
        & = \Delta b^\mathrm{D}(z, k) \\
        & \sim k^{-2},
    \end{split},
\end{equation}
i.e. the general expression of scale-dependent bias in Eq.~(\ref{eq:db_mv}) recovers
the bias proposed in~\citet{2008PhRvD..77l3514D}. The advantage of using Eq.~(\ref{eq:db_mv}) is that it can be
used to calculate any shape of PNG, provided that the bispectrum $B_{\phi}$ function is given.

The scale-dependent bias for equilateral, orthogonal and enfolded shapes of PNG 
can be obtained by substituting Eqs.~(\ref{eq:B-equi}), (\ref{eq:B-orth}) and (\ref{eq:B-enfold})
into Eq.~(\ref{eq:fk}).  In Fig.~\ref{fig:db}, we show the absolute value of the scale-dependent part of the bias, i.e. Eq.~(\ref{eq:Delta-b-z-k}) for the four shapes of PNG at $z=0$ (left panel) and $z=2$ (right panel). One can see that the local shape has the most prominent feastures of scale-dependent bias at large scales, which can be constrained with 21-cm intensity mapping observation on large angular scales. The orthogonal and enfolded shapes have less prominent features but are possibly detectable at small $k$. The scale-dependent bias induced by equilateral shape is too small on large scales so it will be hard to detect. The results shown in Fig.~\ref{fig:db} are consistent with the analysis in~\cite{2015arXiv150705903R} and Fig.~1 in~\cite{Fedeli11}.

We can see the asymptotic behavior of scale-dependent bias [Eq.~(\ref{eq:Delta-b-z-k})] on large scales by taking the limit of $k\rightarrow0$, then $\Delta b\rightarrow(\mathcal{F}/\mathcal{M}_R)$. Therefore,
\begin{equation}
    \begin{split}
        \Delta b (\mathrm{Local})       & \sim k^{-2} \\
        \Delta b (\mathrm{Equilateral}) & \sim {\rm const}  \\
        \Delta b (\mathrm{Enfolded})    & \sim  k^{-1} \\
        \Delta b (\mathrm{Orthogonal})  & \sim  k^{-1}. 
    \end{split}
\end{equation}

These asymptotic behaviors of $\Delta b$ are consistent with the computation of halo models in Fig.~\ref{fig:db}.



\subsection{Power spectrum}
\begin{figure}[htb]
    \centering
    \includegraphics[width=0.5\textwidth]{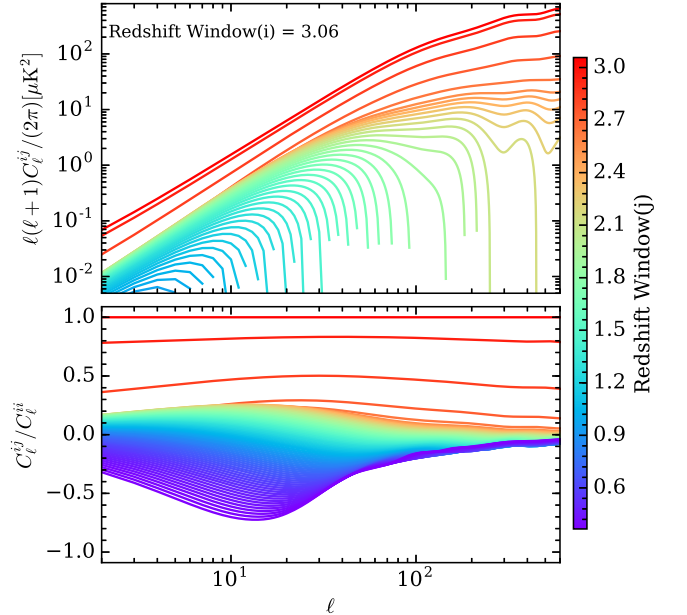}
    \vspace{-0.6cm}
    \caption{
        Upper panel: Cross-correlated angular power spectrum between redshift 
        $z_i=3.06$ and $z_j$, which ranges from $0.37$ to $3.06$ shown with 
        different colors.
        Lower panel: The radio of tomographic angular cross-power spectrum 
        between $z_i$ and $z_j$ to the auto-power spectrum of $z_i$.
    }\label{fig:clij}
\end{figure}

We employ the \hi tomographic angular power spectrum as the 
observable in our analysis,
The expression of the angular power spectrum of the 
$i$th and the $j$th redshift bins is
\begin{equation}
    C_\ell^{ij} = 4\pi T_{\rm b}^{ij} \int \mathrm{d}\,\ln k \mathcal{W}_\ell^i(k) 
    \mathcal{W}_\ell^j(k) \Delta_\zeta^2(k),
\end{equation}
in which, $\Delta_\zeta^2(k)$ is the dimensionless power spectrum of primordial
curvature perturbation and $T_{\rm b}^{ij} = T_{\rm b}(z_i)T_{\rm b}(z_j)$ is the multiplication 
of \hi mean brightness temperature of the $i$th and $j$th redshift bins.
We use the expression of $T_{\rm b}(z)$ in 
\citeauthor{2008PhRvL.100i1303C}(\citeyear{2008PhRvL.100i1303C})
\citep{2008PhRvL.100i1303C}, 
\begin{eqnarray}
        T_{\rm b}(z) &=& 0.39 
        \left(\frac{\Omega_\hi}{10^{-3}}\right)
        \left(\frac{1+z}{2.5}\right)^{0.5}  \nonumber \\
       & \times & 
        \left(\frac{\Omega_{\rm m} + (1+z)^{-3}\Omega_\Lambda}{0.29}\right)^{-0.5}
        \mathrm{mK},
\end{eqnarray}
where $\Omega_\mathrm{\hi}$ is the fractional \hi density assumed to be 
$0.62\times10^{-3}$~\cite{2013MNRAS.434L..46S}.
The window function $\mathcal{W}_\ell(k)$ is,
\begin{equation}
    \mathcal{W}_\ell(k) = \int \mathrm{d}\chi 
    \frac{\mathrm{d}N_g(\chi)}{\mathrm{d}\chi} j_\ell(k\chi)
    b_\mathrm{\hi}^\mathrm{NG}(\chi(z), k)T_\delta(\chi,k),
\end{equation}
where $j_\ell$ is a spherical Bessel function, 
$\mathrm{d}N_g(\chi)/\mathrm{d}\chi$ is the redshift 
distribution of galaxy number, $T_\delta(\chi, k)$ is the transfer 
function for the galaxy number over-density, and $b_\mathrm{\hi}^\mathrm{NG}$
is the total bias of \hi (Eq.~(\ref{eq:bNG})). To calculate the angular power
spectrum, we use the \cambs package~\cite{2011PhRvD..84d3516C}.

Figure~\ref{fig:clij} shows the tomographic angular power spectrum.
The upper panel shows the cross-power spectrum between redshift
$z_i=3.06$ and $z_j$, which ranges from $0.37$ to $3.06$ shown with
different colors. The lower panel shows the ratio of
the cross-power spectrum of different redshift bins to the 
auto-power spectrum of the same redshift bin.
We can see that the cross-power spectrum decreases as the 
redshift deviates from $z_{i}=3.06$. This is what we expected, since the cross-correlated
signal should drop if the frequency windows move away from each other.

\section{fisher matrix forecast}\label{sec:fisher}

To forecast the potential for constraining $f_\mathrm{NL}$, 
we perform the Fisher matrix analysis. If we assume that the 
model likelihood surface in parameter space can be well
approximated by a multivariant Gaussian, the Fisher matrix
$\bf{F}$ is then a good approximation for the inverse of the 
parameter covariance. In the 21-cm tomography, each frequency band will provide a map of 21-cm intensities,
so we need to sum over the Fisher matrix in both $\ell$-space and frequency space. Since $\nu=1420{\rm MHz}/(1+z)$, each frequency corresponds to a unique redshift slice. The Fisher matrix is
\begin{equation}
    {\bf F}_{\alpha\beta} = f_\mathrm{sky} 
    \sum^{\ell_\mathrm{max}}_{\ell_\mathrm{min}}
    \left(\frac{2\ell+1}{2}\right) \mathrm{tr}[{\bf C}_{\ell,\alpha} {\bf \Sigma}_\ell 
    {\bf C}_{\ell,\beta} {\bf \Sigma}_\ell],
\end{equation}
in which ${\bf C}_\ell$ is an $n_z \times n_z$ matrix, in which each element is
the \hi cross angular power spectrum between the two frequency bins. ${\bf \Sigma}_\ell = ({\bf C}_\ell + {\bf N}_\ell)^{-1}$ is the {\it total} noise
inverse matrix, in which
${\bf N}_\ell$ is the $n_{z} \times n_z$ experimental noise power spectrum. Here we make a simple assumption that
the noises in different frequency (redshift) bins are uncorrelated, therefore the ${\bf N}_\ell$ is a diagonal matrix. In reality, 21-cm intensity maps are highly contaminated by the foreground, such as Galactic synchrotron emission, extragalactic point sources, and atmospheric signal. One needs to apply foreground removal technique to reduce the foreground contamination~\cite{Zhang16,Bigot15,Olivari16}. However, there always be some level of residual Galactic foreground after applying such techniques to the maps. Therefore the cross-correlation of noises between different frequency bands may not completely be zero. 

Under our simplified assumption, the element of ${\bf N}_\ell$ matrix is
\begin{eqnarray}
    N_\ell^{ij} &=& \delta^{ij}N_\ell^\hi  \nonumber\\
                &=& \delta^{ij}T_\mathrm{sys}^2 
                S_\mathrm{survey}/
                (N_\mathrm{ant}N_\mathrm{feed}t_\mathrm{TOT}\Delta\nu).
\end{eqnarray}
$T_\mathrm{sys}=T_{\rm rec}+T_{\rm sky}$ is the system temperature, which is contributed from
the sky temperature,
$T_\mathrm{sky} = 60 \times (300\mathrm{MHz}/\nu)^{2.55}$, and 
receiver temperature $T_\mathrm{\rm rec}$ for each experiment. $N_{\rm ant}$ and $N_{\rm feed}$ are the number of antenna and the number of feed horn in each antenna respectively. The detailed experimental parameters
for FAST, SKA-I and BINGO are listed in Table~\ref{tab:fast}.

\section{Experiment parameters}\label{sec:exp}

\begin{figure*}[htb]
    \centerline{
    \includegraphics[width=0.5\textwidth]{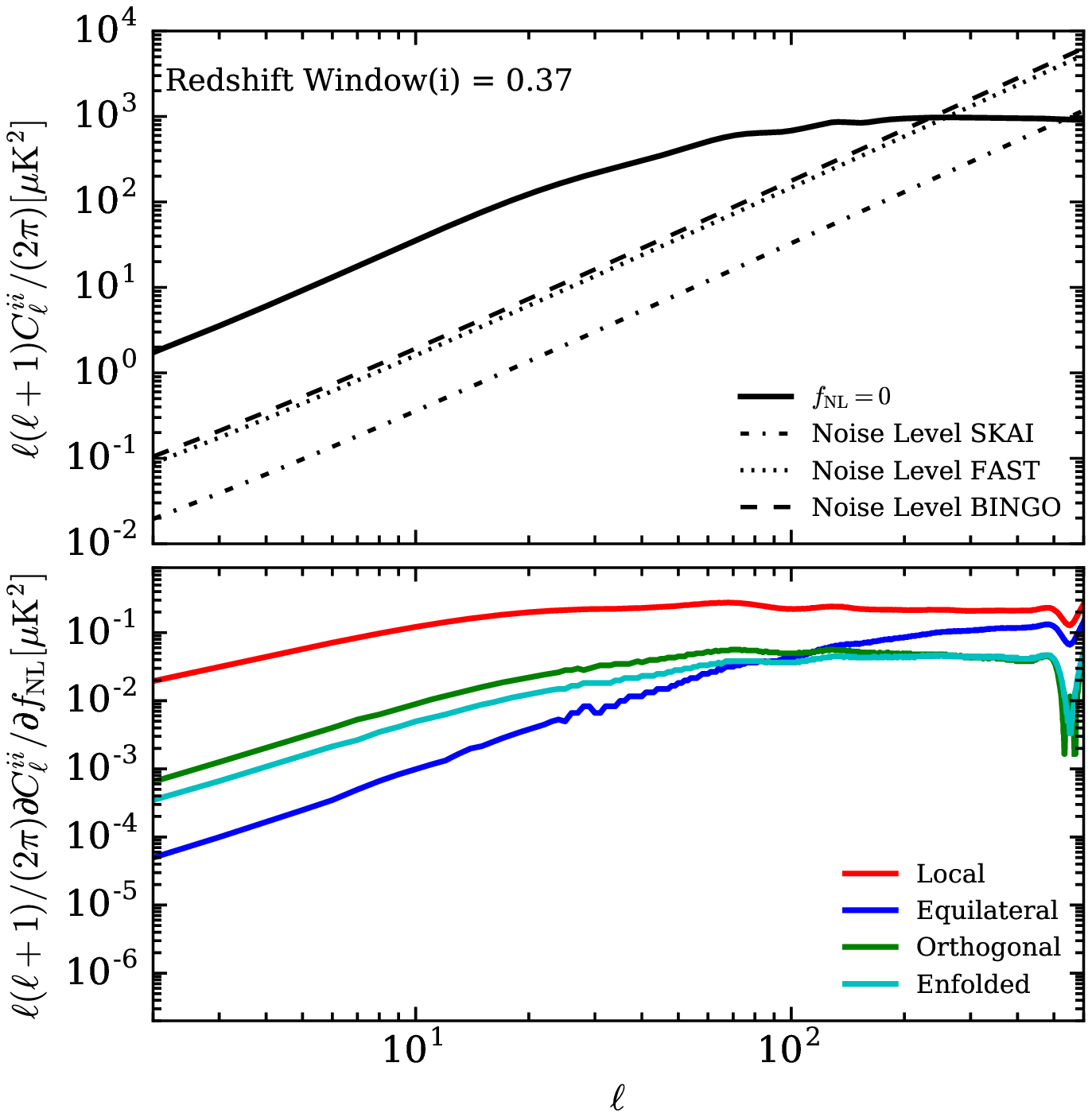}
    \includegraphics[width=0.5\textwidth]{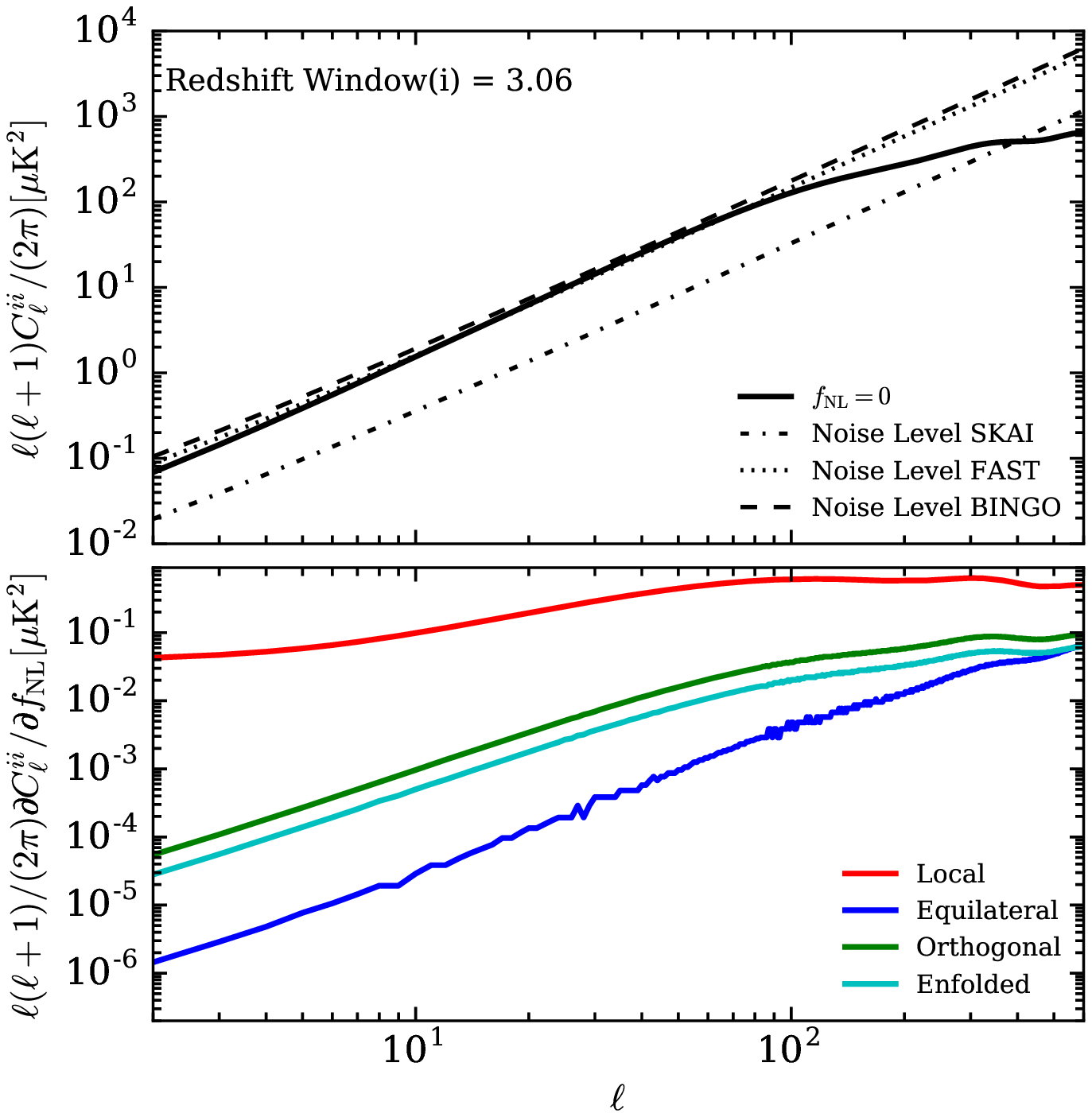}}
    \vspace{-0.6cm}
    \caption{Upper panels: Comparison between the noise power 
    spectra of different experiments and the 21-cm power spectrum in 
    standard model ($f_{\rm NL}=0$) for the two representative redshift bins 
    (left and right panels). In both panels, one-year observation time (equivalent to $3.15\times 10^{7}$sec) 
    and $2500\deg^2$ survey area are assumed for all the experiments.  
    Lower panels: The partial derivatives of $C_\ell^{ii}$ with respect 
        to parameter $f_\mathrm{NL}$ for four shapes of PNG.}\label{fig:clii}
\end{figure*}

\begin{table}[hb]
    \small
    \centering
    \caption{
        The experiment parameters for FAST, SKA-I and BINGO.
        $D_\mathrm{dish}$ is the illuminated aperture.
    }\label{tab:fast}
\begin{tabular}{L{2cm}|C{1.5cm}C{1.5cm}C{1.5cm}} \hline\hline
                                      & FAST        & SKA-I       & BINGO\\\hline
$\nu_\mathrm{min}[\mathrm{MHz}]$      & $1050$      & $350$       & $960$\\
$\nu_\mathrm{max}[\mathrm{MHz}]$      & $1350$      & $1050$      & $1260$\\
$\Delta\nu[\mathrm{MHz}]$             & $10$        & $10$        & $10$\\
$n_\nu (n_z)$                         & $30 $       & $70$        & $30$\\
$D_\mathrm{dish}[\mathrm{m}]$         & $300$       & $15$        & $25$\\
$N_\mathrm{ant}\times N_\mathrm{feed}$& $1\times19$ & $190\times1$& $1\times60$\\
$t_\mathrm{TOT}[\mathrm{yr}]$         & $1$         & $1$         & $1$\\
$T_\mathrm{rec}[\mathrm{K}]$    & $25$        & $28$        & $50$\\
$S_\mathrm{survey}[\mathrm{deg}^2]$   & $<24000$    & $<25000$    & $2500$\\\hline\hline

\end{tabular}
\end{table}

\paragraph{BINGO} 
The BINGO experiment is a single-dish \hi intensity 
mapping experiment, which aims at mapping the \hi emission at frequencies 
between $960 \mathrm{MHz}$ and $1260\mathrm{MHz}$~\cite{2013MNRAS.434.1239B,Battye16}. 
The telescope of the BINGO experiment has no moving parts and it conducts a drift-scan strategy. 
To achieve enough 
survey area, a wide instantaneous field of view (FOV) with multiple feeds
is required. A total of 60 feeds laid out in a rectangle of 
$16\mathrm{m}\times 15\mathrm{m}$ at the focal plane. This will form 
a FOV of about $10^\circ$(in Declination direction)
$\times 9^\circ$(in Right Ascension direction). With the $10^\circ$
wide strip centering at Declination of $-45^\circ$, the total survey
area is about $2500\deg^2$.

\paragraph{FAST}
FAST is the largest single-dish telescope, which also has the 
multibeam system of 19 feed-horns array~\cite{2011IJMPD..20..989N, 
2016RaSc...51.1060L}. 
The multibeam system is proposed to work at frequencies from 
$1.05$ to $1.45\mathrm{GHz}$ with system temperature 
of $25\mathrm{K}$. In our analysis, we only include the frequencies 
up to $1.35\mathrm{GHz}$. With the $300\mathrm{m}$ illuminated aperture,
each of the feed-horn has the beam size (Full Width at Half Maximum) 
of $2.9'$, and form a $26'$ FOV with $19$ beams.
Due to the long slewing time, FAST can only work on drift-scan 
observation mode. Similar to the BINGO experiment, FAST scans 
a $26'$ wide strip along the Right Ascension direction for each sidereal day.
But the zenith angle of FAST can be adjusted from Dec:$-14^\circ12'$ to
Dec:$65^\circ48'$. Without over lapping between scanning strips, 
it takes about half year to cover all $80^\circ$ Declination range . 
With one-year observation ($3.15 \times 10^{7}$ second), the maximum survey area is about $24000\deg^2$.

\paragraph{SKA-I} The SKA Phase I (SKA-I) plans to construct $190$
movable $15\mathrm{m}$ dishes~\cite{Bull15}. The maximum survey area is about 
$25000\deg^2$. A efficient survey area is need to be explored to 
minimal the constraint errors. In our analysis, we only consider the 
autocorrelation of each dishes, which means that the SKA-I works as 
$190$ single dishes. Without the interferometry, the SKA-I has
very low resolution and is only sensitive to the low-$\ell$ modes.

Figure~\ref{fig:clii} shows the noise power spectra of different experiments 
in at redshift bin $z=0.37$ (left upper panel) and $z=3.06$ (right upper panel). 
The black solid line in the upper panel of each figure shows the standard angular
power spectrum of 21-cm ($f_{\rm NL}=0$); The black dash-dotted, dotted and dashed lines
show the noise power spectra of SKA-I, FAST and BINGO experiments.
One-year observation time and $2500\deg^2$ survey area are assumed 
for all the experiments.
The partial derivatives of $C_\ell^{ii}$ with respect to parameter 
$f_\mathrm{NL}$ are shown in the lower panel. The different colors
correspond to different types of PNG.

Comparing to the BINBO experiment, FAST and SKA-I can have very large survey area.
However, with the limit integration time, the large survey area
may not be able to beat down the constraint error.
We will discuss the details in Sec.~\ref{sec:result}.

\section{results and discussion}\label{sec:result}
\begin{figure*}[htb]
    \centerline{
    \includegraphics[width=0.34\textwidth]{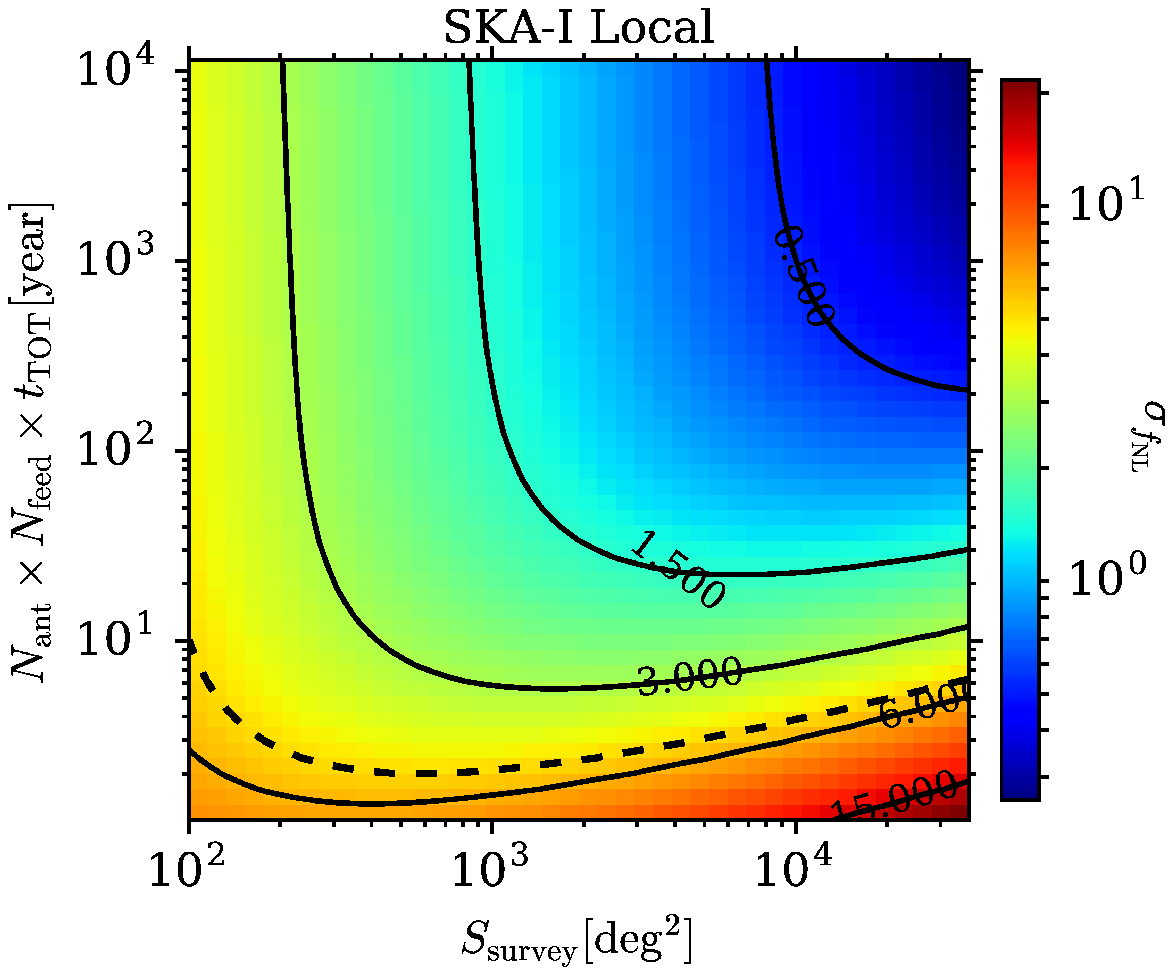}
    \includegraphics[width=0.34\textwidth]{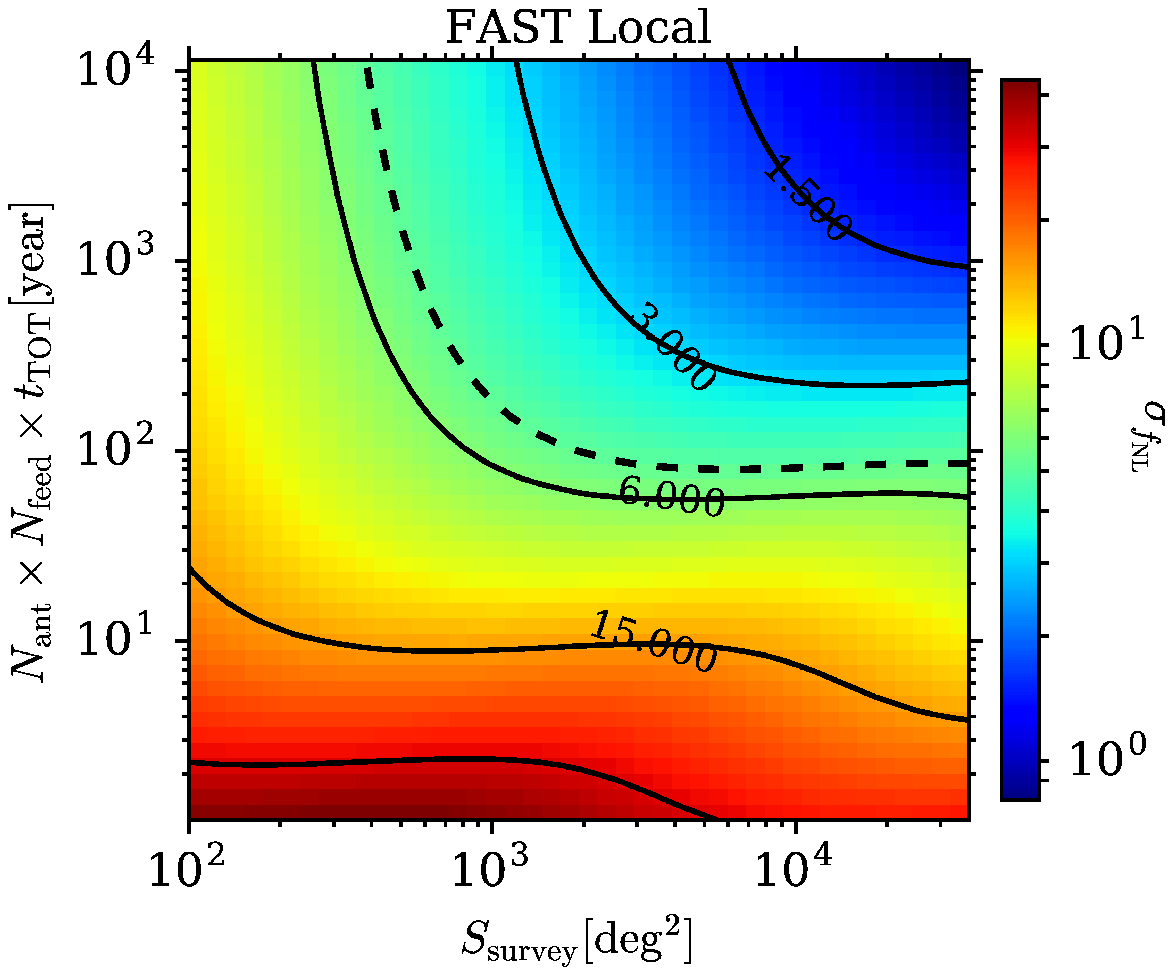}
    \includegraphics[width=0.34\textwidth]{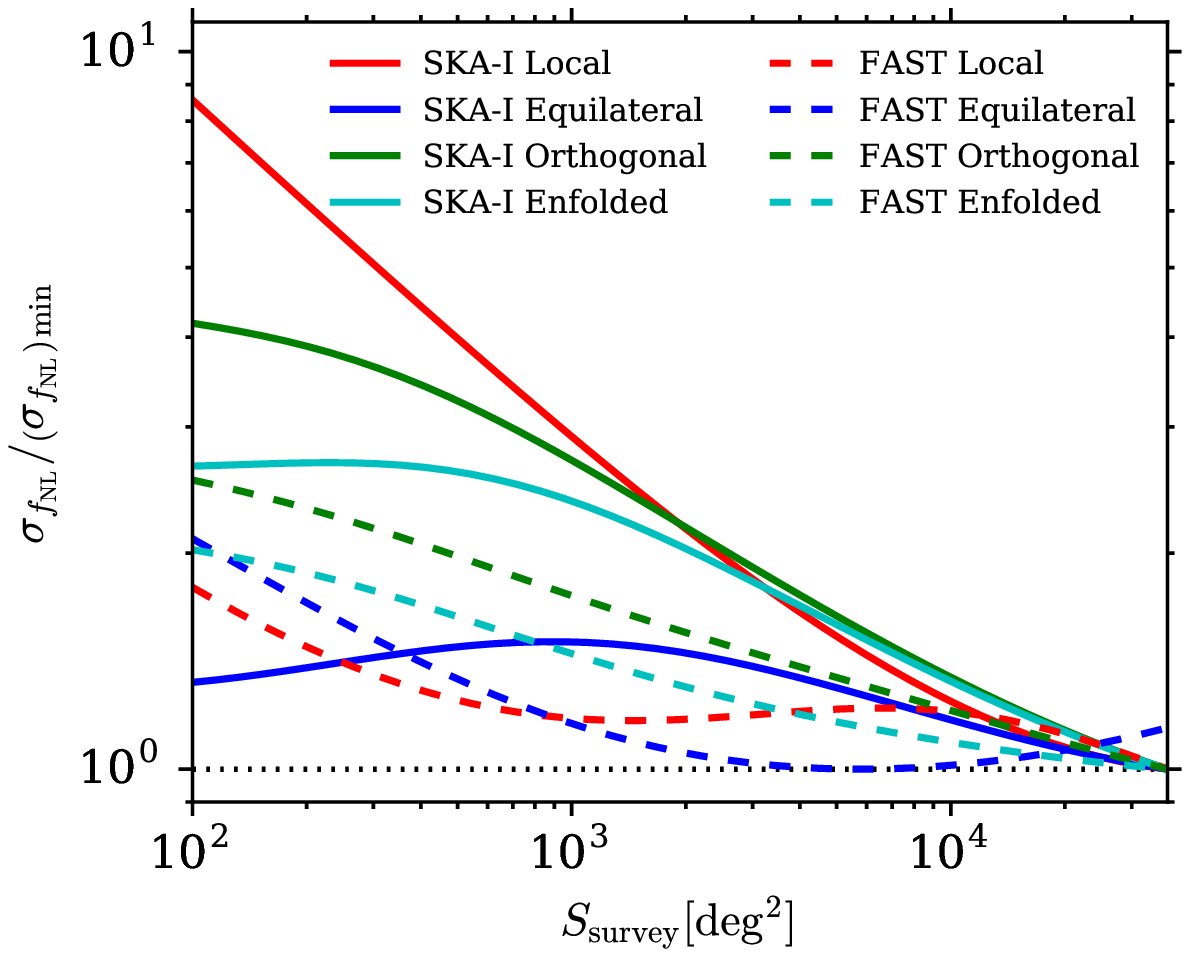}}
    \vspace{-0.3cm}
    \caption{
        The left (for SKA-I) and middle (for FAST) panels show the $\sigma_{f_\mathrm{NL}}$ contours 
        for local-shape PNG in the parameter space of the 
        survey area and total observation time. The dashed contour
        is the error of constraint with {\it Planck} temperature and 
        polarization data~\cite{2016A&A...594A..17P}. The right panel shows the 
        $\sigma_{f_\mathrm{NL}}/(\sigma_{f_\mathrm{NL}})_\mathrm{min}$ 
        as a function of survey area for
        various PNG types. The solid lines
        show the results for SKA-I with one-year observational time and
        $190$ dishes; the dashed lines show the results for FAST with
        one-year observational time and $19$ beams.
    }\label{fig:ska_ts}
\end{figure*}

Figure~\ref{fig:ska_ts} shows the $\sigma_{f_\mathrm{NL}}$ contours for 
local-shape PNG in the plane of the 
survey area and total observation time.
The left and middle panels of Fig.~\ref{fig:ska_ts} show the contours for SKA-I and FAST experiments respectively.
The color going from red to blue means that the constraints become stronger. Different black solid lines are the contours of the same error of $f^{\rm local}_{\rm NL}$. Therefore, the error tends to become smaller if $N_{\rm ant} \times N_{\rm feed} \times t_{\rm TOT}$ becomes bigger. Thus the most efficient way to reduce the constraint 
error is to increase the observation time or the number of dishes(feeds).
Assuming one-year observation time and the maximum dish(feeds) number
for SKA-I and FAST experiments, the constraint errors of
various PNG types as a function of survey area 
are shown in the right panel of Fig.~\ref{fig:ska_ts}.
In order to have a clear view, the constraint errors, $\sigma_{f_\mathrm{NL}}$,
are divided by the their minimal values.
It is true that the optimal survey area may not be the maximal survey
area. For example, 
in the case of equilateral shape, the optimization is 
about $6000\deg^2$ for the FAST experiment.
For other shapes, the optimized survey areas are approaching the maximum sky coverage of SKA-I or FAST.
The large survey area can help to beat the cosmic variance on
large scales, but the integration time per pixel becomes smaller, leading to larger pixel noise. 

One can see from the right panel of Fig.~\ref{fig:ska_ts} that, generally speaking, the larger the survey area is, the smaller the error of $f_{\rm NL}$, except for measuring equilateral shape of PNG using the FAST survey. This is different from the situation of using 21-cm intensity mapping to measure the angular scale of BAO acoustic oscillation, which have the optimal survey area around $6000\,$deg$^{2}$ (For BINGO, see Fig.~7 in~\cite{2013MNRAS.434.1239B}, and for FAST, see Fig.~1 in~\cite{Bigot16a}). The reason is because scale-dependent bias from PNG is always prominent on very large scales, so beating down cosmic variance is more important than lowering down the pixel noise. However, BAO scale is subhorizon for which there is always a trade-off between lowering down pixel noise and beating down cosmic variance. We use different optimized survey areas for different cases 
in the later analysis.

\begin{figure*}[htb]
    \centerline{
    \includegraphics[width=0.45\textwidth]{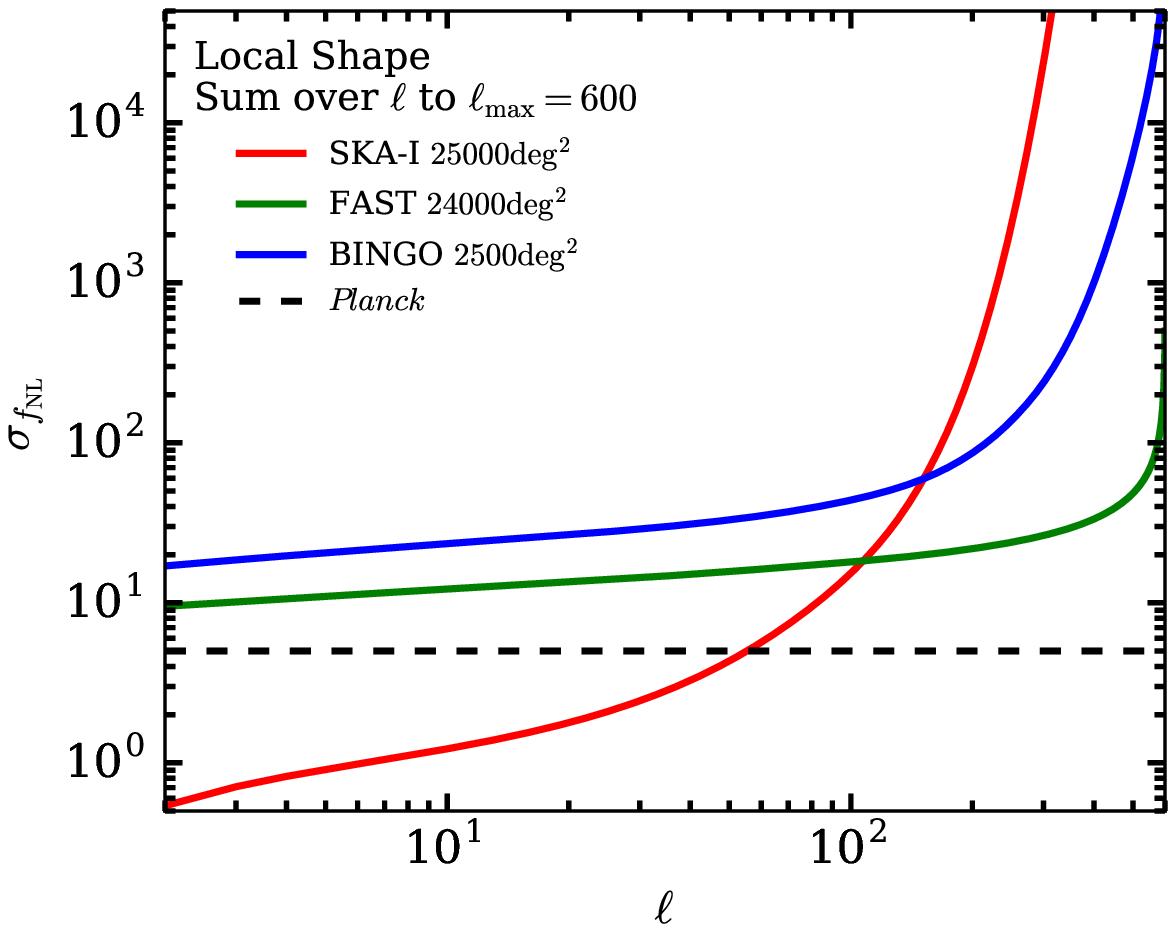}
    \includegraphics[width=0.45\textwidth]{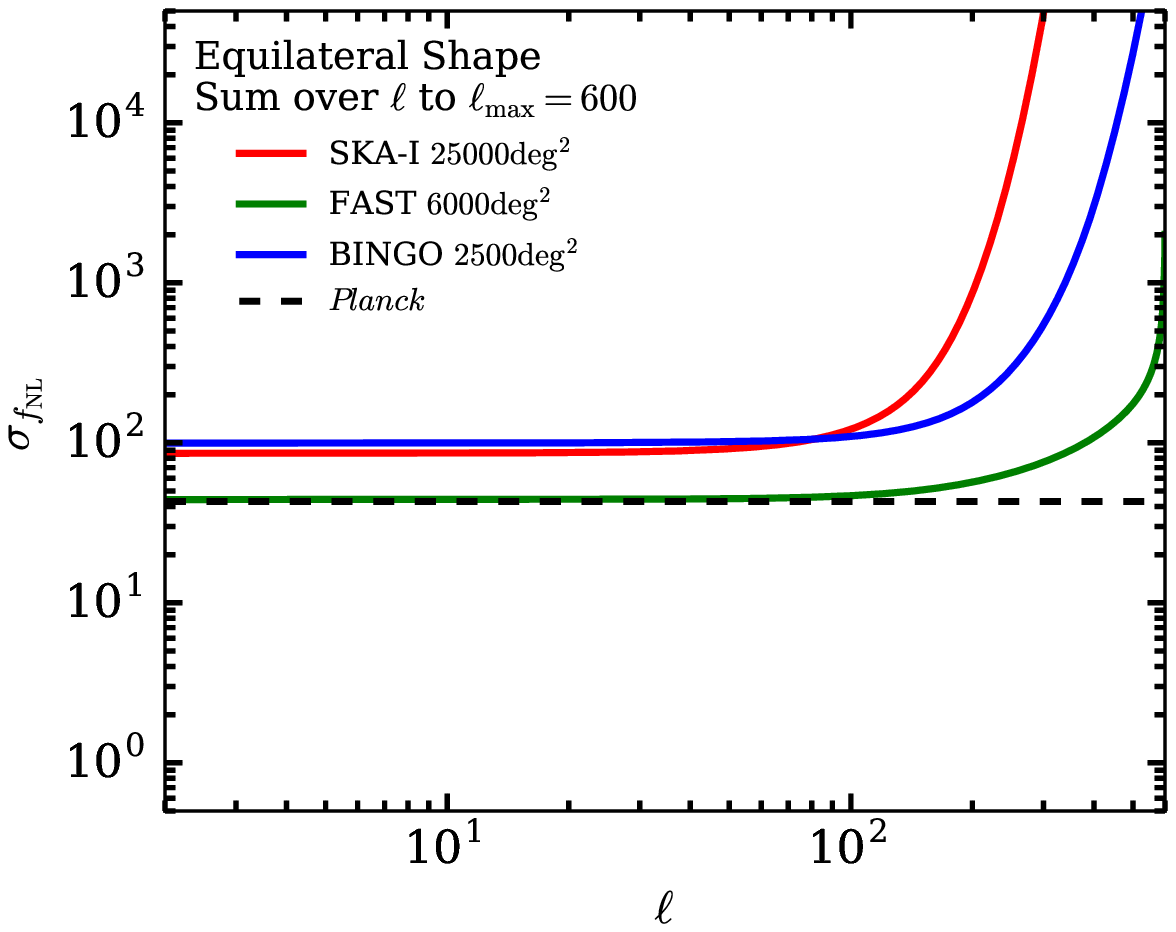}}
    \centerline{
    \includegraphics[width=0.45\textwidth]{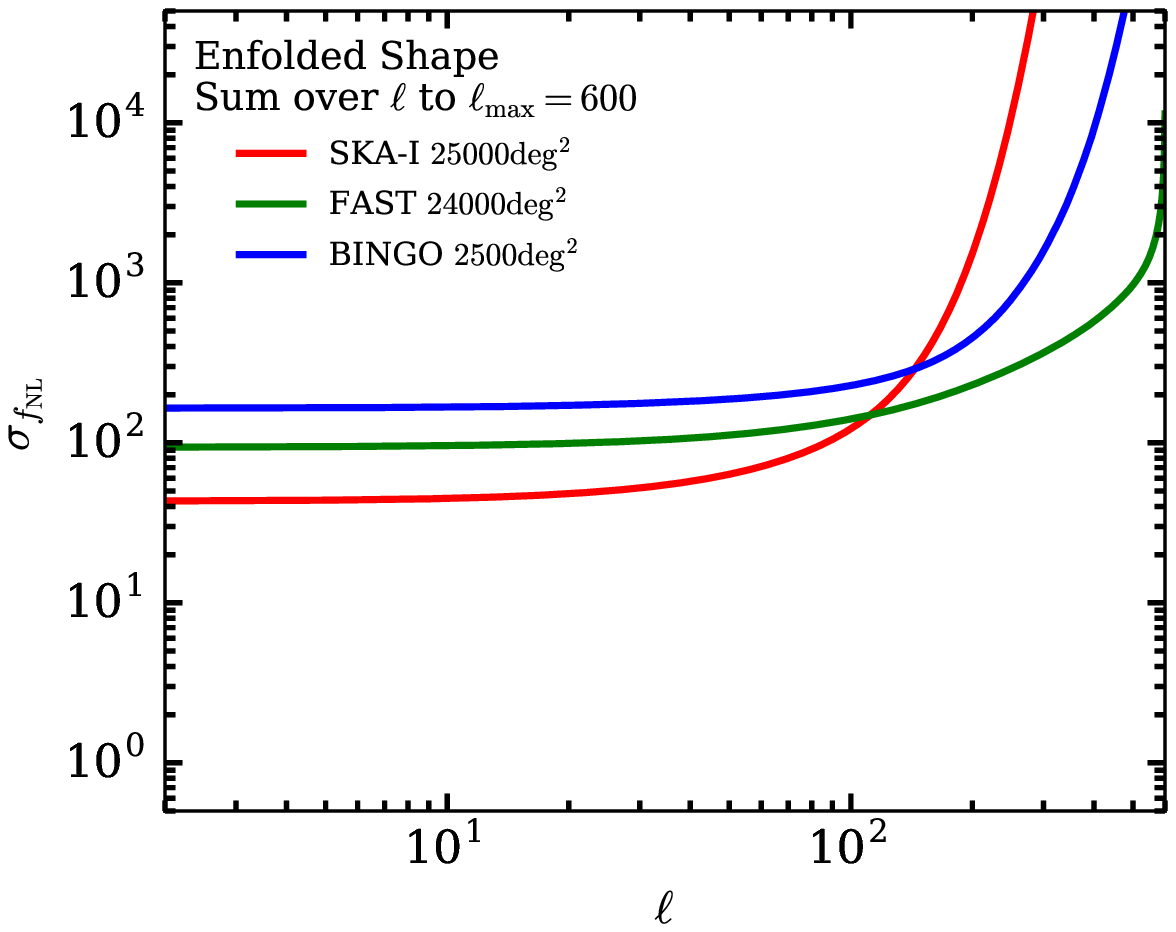}
    \includegraphics[width=0.45\textwidth]{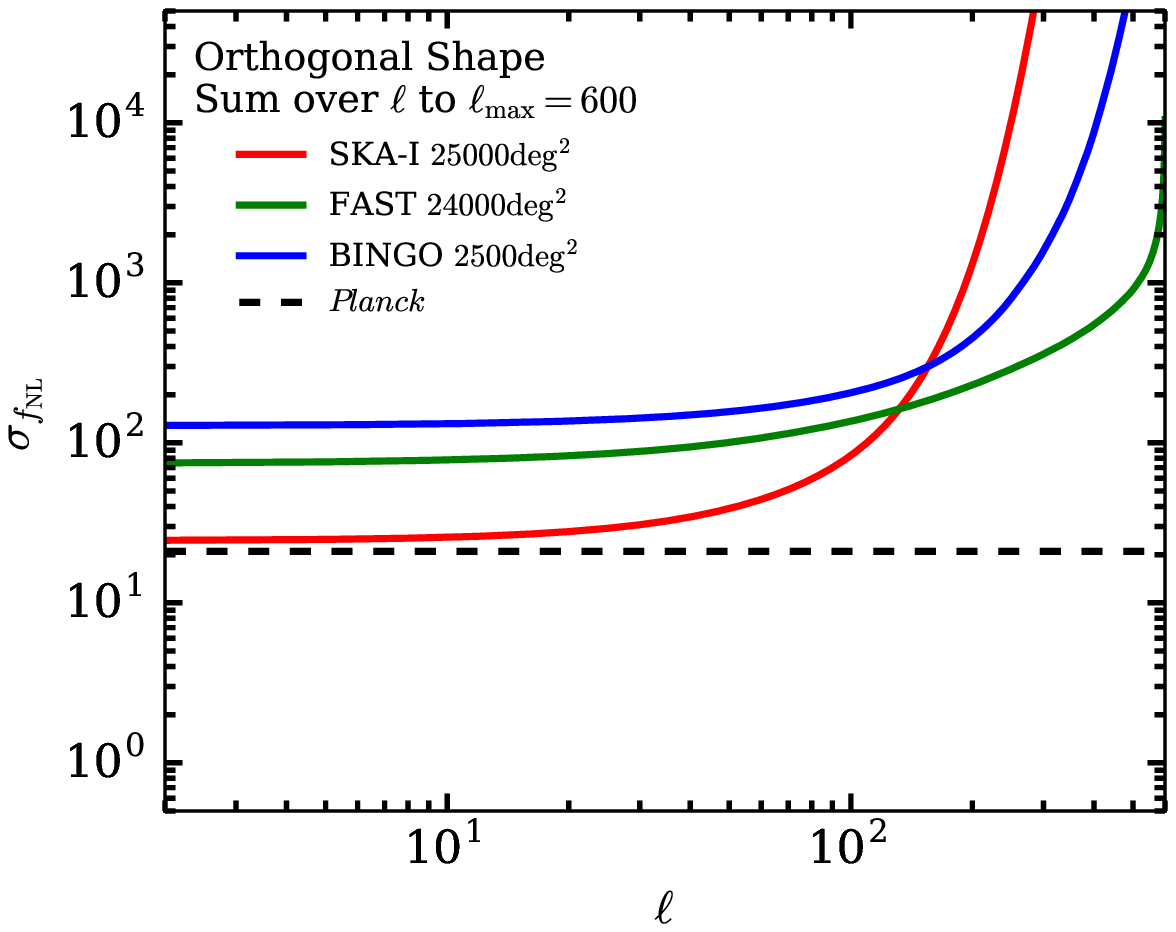}}
    \vspace{-0.6cm}
    \caption{
        The $\sigma_{f_\mathrm{NL}}$ as a function of $\ell_\mathrm{min}$
        for various experiments and PNG shapes.
        The black dashed line is the current constraint with 
        {\it Planck} temperature and polarization data~\cite{2016A&A...594A..17P}.
    }\label{fig:fisher}
\end{figure*}

Figure~\ref{fig:fisher} shows the $\sigma_{f_\mathrm{NL}}$ as a function
of $\ell_\mathrm{min}$ if we fix $\ell_{\rm max}=600$. Different PNG shapes are
shown in different panels. In each panel, different colors indicate
different experiments as shown in the legends. The optimized survey
areas are applied to the analysis. The constraint errors of different PNG shapes from
{\it Planck} satellite are 
shown with the black dashed lines~\cite{2016A&A...594A..17P}. 
The $\sigma_{f_\mathrm{NL}}$ of different PNG 
shapes forecasted with different experiments are listed in Table~\ref{tab:result}.

\begin{table*}[htb]
    \small
    \centering
    \caption{
        $\sigma_{f_\mathrm{NL}}$ of different PNG 
        shapes forecasted with different experiments. The optimized 
        survey areas are applied in the analysis.
        The ``{\it Planck} 2015" column shows the constraint error with 
        Plank temperature and polarization data~\cite{2016A&A...594A..17P}.
The numbers in bold character are the constraints better than {\it Planck}. 
    }\label{tab:result}
\begin{tabular}{L{1.7cm}|C{1.7cm}|C{1.5cm}C{1.5cm}C{1.5cm}|C{1.6cm}C{1.6cm}C{1.6cm}} \hline\hline
                &                  & \multicolumn{3}{c|}{ Current Configuration} & \multicolumn{3}{c}{ Extentions}  \\
                & {\it Planck} 2015& FAST & SKA-I & BINGO & SKA-I 2yr$^{\dagger}$  & FAST 2yr$^{\dagger\dagger}$    & FAST low$^{\ddagger}$ \\ \hline
    Local       & 5   & 9.5  & {\bf 0.54} &  17   & {\bf 0.43} & 7.4      & {\bf 1.6} \\
    Equilateral & 43  & 44   & 86         & 100   & 66         & {\bf 32} & 53  \\
    Orthogonal  & 21  & 75   & 25         & 128   & {\bf 20}   & 59       & 39  \\
    Enfolded    & --  & 94   & 43         & 164   & 36         & 70       & 64  \\ \hline\hline
    \multicolumn{8}{L{12.5cm}}{
        {\footnotesize
        $^{\dagger}$ SKA-I with two-year observation;
        $^{\dagger\dagger}$ FAST with two-year observation;
        $^{\ddagger}$ FAST with low frequencies range from $350\mathrm{MHz}$ to $1050\mathrm{MHz}$}
    }
\end{tabular}
\end{table*}

We can see that, for the local shape PNG, 
the SKA-I experiment is potentially able to constrain $f_\mathrm{NL}$ better 
than {\it Planck} experiment. But we should realize that it is only the most
ideal case. It is well known that, one of the big challenges 
for observations of \hi intensity mapping is the foreground subtraction, and the low-$\ell$
modes may not be detectable due to the foreground contamination.
Our results show that, to obtain a remarkable constraint on $f_\mathrm{NL}$ 
with the SKA-I intensity mapping in the future, we need to recover 
the angular power spectrum of \hi with the minimal $\ell_{\rm max}\simeq 50$. This is the aim of
several recent efforts of restoring large angular power with cross-correlation with weak gravitational lensing~\cite{Zhu16a,Zhu16b}. We also find that the constraint error for orthogonal shape 
PNG with SKA-I is $\sim25$, which is
at the same level of current {\it Planck} limit. If the observation 
can be extended to $2$ years, the error will be reduced to $\sim20$.

The constraint error for equilateral shape
PNG with FAST is $\sim44$, which 
is better than the results of SKA-I and BINGO experiments. 
The FAST error on $f^{\rm equil}_{\rm NL}$ 
is close to the current limit of {\it Planck} experiment. 
This is because the scale-dependent bias induced by the equilateral shape 
PNG has higher signal-to-noise ratio at small
scales and the FAST experiment is more sensitive to the small-scale
modes than SKA-I single dish mode and BINGO. 
So far, in our analysis, we assume perfect knowledge of the power spectrum
and do not include the theoretical error.
However, it has been shown that the higher derivative terms contribute 
to the scale-dependent bias on small scales
\cite{2016arXiv160200674B, 2015JCAP...12..043A}. 
Such contributions induce extra uncertainties to the scale-dependent bias
measurements and reduce the detectability of equilateral PNG.


We also test the possible extensions of the current configuration by adding more integration time.
If the observation time for SKA-I and FAST could be extended to $2$ years, the constraints on $f_{\rm NL}$
can be improved quantitatively. The forecasted constraint on different shapes of PNG are listed
in Table~\ref{tab:result}. It is worth noticing that the constraint
error on the orthogonal-shaped PNG with SKA-I and the equilateral-shaped PNG
with FAST becomes smaller than the limits of {\it Planck} with extended observational time.

A good extension for FAST experiment is to extend its bandwidth 
to the lower frequencies, which are corresponding to the higher
redshifts.  
So far the FAST telescope has one ultrawide band receiver working on 
$270\mathrm{MHz}\sim1.62\mathrm{GHz}$. Unfortunately, the ultrawide
band receiver has only one beam. It will take quite a long time to achieve
the same observation time as the multibeam receiver.
Now the multibeam system of the FAST telescope is designed to
work on frequencies between $1050\mathrm{MHz}$ and $1350\mathrm{MHz}$.
Assuming that the FAST multibeam system works on the frequencies between 
$350\mathrm{MHz}$ and $1050\mathrm{MHz}$, which is the same as 
the frequency range of SKA-I experiment, the constraint for local shape 
PNG will be $\sigma_{f^{\rm local}_\mathrm{NL}}\sim1.62$ with the 
optimized survey area of $6000\deg^2$. The constraint errors ($\sigma_{f_{\rm NL}}$) for
orthogonal and enfolded shapes become $39$ and $64$ respectively, which are all highly reduced.

\section{conclusion}
In this work, we explored the constraining power on the 
primordial non-Gaussianity (PNG), with the future single-dish \hi intensity mapping 
observations with BINGO, FAST and SKA-I.
Four fundamental shapes of PNG are studied in our analysis, 
including local, equilateral, orthogonal and enfolded. 
We focus on the effect of scale-dependent bias to the underlying 
dark matter tracer, induced by the primordial non-Gaussinaity.
The properties of such scale-dependent bias at large-scale limit 
are discussed in our analysis.  
The forecast results are listed in Table~\ref{tab:result}.

Our forecasts show that with the current configuration of the experiments
one-year observation time, the constraint on local shape of PNG from SKA-I intensity mapping experiment can be 
better than the current {\it Planck} experiment.
The optimized survey area of $25000\deg^2$ is applied in the analysis of SKA-I,
but the results are more sensitive to the total observation time than 
the survey area. However, the \hi intensity mapping experiments may be
contaminated by the foreground and the low-$\ell$ modes may be be
detectable. Our analysis shows that the SKA-I experiment can still 
have the remarkable constraint without the modes of $\ell\lesssim50$
With two-years observation, the constraint on orthogonal shape
PNG is $\sim20$, which is also better than
the constraint from {\it Planck} measurement.

The FAST experiment has the advantage of higher angular resolution
and is more sensitive to the small-scale modes, which is good for 
constraining the equilateral shape of PNG.
With the current configuration and two years observation, the 
constraint error for the equilateral shape of PNG 
will be $\sigma_{f_\mathrm{NL}}=32$, which is better than the current limit of the
{\it Planck} observation. However, such a limit is achieved by ignoring 
the extra uncertainties caused by the higher derivative terms. 
Previous studies show that such extra uncertainties may not be negligible.
The detailed limit for the equilateral-type PNG needs to be investigated in
the future analysis.

Similar constraint on the local shape of PNG 
can be achieved by the FAST \hi intensity mapping, if its frequency
bandwidth can be extended to the lower frequencies (ultrawide band). Assuming the
same working frequency range, the best constraint from FAST on 
the local shape of primordial non-Gaussianiy is $\sigma_{f_\mathrm{NL}}\sim1.62$.

The studies we conduct here are the standard power spectra analysis of 21 cm. There have been efforts on 
using the multitracer technique to beat the cosmic variance and obtain tighter constraints on $f_{\rm NL}$~\cite{Abramo13,Fonseca15,Alonso15}. In addition, using three-point correlation function is another way to measure PNG. These methods will be explored to measure all shapes of $f_{\rm NL}$ in the future work.

\acknowledgements
We thank Neal Dalal, Di Li, 
Roy Maartens, Jerome Gleyzes, Yi Wang and Xiao-Dong Xu for helpful discussions and 
Stefano Camera for his help on \cambs. This work is 
supported by the National Research Foundation of South Africa with 
Grant no.105925 and the University of KwaZulu-Natal staff start-up grant.


\bibliography{draft_v9}

\end{document}